# Facile Synthesis of High Quality Graphene Nanoribbons

Liying Jiao, Xinran Wang, Georgi Diankov, Hailiang Wang & Hongjie Dai*

*Department of Chemistry and Laboratory for Advanced Materials, Stanford University, Stanford, California 94305, USA*

* Correspondence to hdai@stanford.edu

**Graphene nanoribbons have attracted attention for their novel electronic and spin transport properties[1-6], and because nanoribbons less than 10 nm wide have a band gap that can be used to make field effect transistors[1-3]. However, producing nanoribbons of very high quality, or in high volumes, remains a challenge[1, 4-18]. Here, we show that pristine few-layer nanoribbons can be produced by unzipping mildly gas-phase oxidized multiwalled carbon nanotube using mechanical sonication in an organic solvent. The nanoribbons exhibit very high quality, with smooth edges (as seen by high-resolution transmission electron microscopy), low ratios of disorder to graphitic Raman bands, and the highest electrical conductance and mobility reported to date (up to $5e^2/h$ and 1500 $cm^2$/Vs for ribbons 10-20 nm in width). Further, at low temperature, the nanoribbons exhibit phase coherent transport and Fabry-Perot interference, suggesting minimal defects and edge roughness. The yield of nanoribbons was ~2% of the starting raw nanotube soot material, which was significantly higher than previous methods capable of producing high quality narrow nanoribbons[1].**

**The relatively high yield synthesis of pristine graphene nanoribbons will make these materials easily accessible for a wide range of fundamental and practical applications.**

Lithographic[4,5,7], chemical[8-11] and sonochemical[1,12] methods have been developed to make graphene nanoribbons. Recently, nanoribbon formations by unzipping carbon nanotubes were reported[13-18]. Two groups successfully unzipped chemical vapor deposition (CVD)-grown multiwalled carbon nanotubes in solution-phase by using potassium permanganate oxidation[14] and lithium and ammonia reactions[16], respectively. Only heavily oxidized and defective nanoribbons were made due to extensive oxidation involved in the unzipping process. We developed an approach to high quality narrow nanoribbons by unzipping nanotubes using a masked gas-phase plasma etching approach[13]. However, the method was limited to nanoribbons formation on substrates. More recently, unzipping methods such as catalytic cutting[17] and high current pulse burning[18] have been reported, but the quality and yield of nanoribbons were unknown. Thus far, a method capable of producing large amounts of high quality nanoribbons is still lacking.

Here we present a new method to unzip nanotubes by a simple two-step process (Fig. 1A). First, raw soot materials containing pristine multiwalled carbon nanotubes synthesized by arc discharge (Bucky tube, Aldrich) were calcined in air at 500 °C. This was a mild condition known to remove impurities and etch/oxidize multiwalled carbon nanotubes at defect sites and ends without oxidizing pristine sidewalls of nanotubes[19]. Then, nanotubes were dispersed in a 1,2-dichloroethane (DCE) organic solution of poly(m-phenylenevinylene-co-2,5-dioctoxy-p-phenylenevinylene) (PmPV) by sonication. During sonication, the calcined nanotubes were found to unzip into nanoribbons with high efficiency. Ultracentrifuge was then used to remove the remaining nanotubes, resulting in high percentage (> 60%) of nanoribbons in the supernatant (Supplementary Fig S1). The yield of nanoribbons was estimated to be ~2 % of the starting raw soot material through the two step process, which could be further improved by repeating the unzipping process for remaining nanotubes in the centrifuged aggregate, increasing the calcination temperature and prolonging the sonication time. The yield and quantity of high quality nanoribbons (width 10-30 nm) far exceeds previous methods capable of making high quality nanoribbons[1,13].

We used atomic force microscope (AFM) to characterize multiwalled carbon nanotubes and the unzipped products deposited on $SiO_2$/Si substrates. Nanoribbons were easily distinguished from multiwalled carbon nanotubes due to obvious decreases in apparent heights (1-2.5 nm in height for nanoribbons, Fig. 1 B to D, Supplementary Fig. S2). The average diameter (height) of the starting nanotubes was ~8 nm. We observed from topographic heights of the nanoribbons (1-2.5nm including PmPV on both sides of the ribbons) that most of the ribbons were either single-, bi- or tri-layered with widths in 10-30 nm range (Fig. 2A and B, Supplementary Fig. S2 and S3). Under AFM, the nanoribbons appeared very uniform in width with little edge roughness along their lengths (Fig. 2B and Supplementary Fig. S3). The high yield of nanoribbons enabled us to readily characterize nanoribbons by TEM (which was not the case in Ref.1). We observed a ~12 nm wide nanoribbons with a fold along its length by TEM (Fig. 2C). The kink structure (Fig. 2D) illustrated excellent flexibility of nanoribbons compared to rigid multiwalled carbon nanotubes. High resolution TEM of our nanoribbons revealed straight and nearly atomically smooth edges without any discernable edge roughness (Fig. 2E and F, Supplementary Fig. S4). This is the first time nearly atomically smooth edges of narrow (< 20 nm) nanoribbons are observed in TEM. The parallel lines seen at the edges of the nanoribbon (Fig. 2E and F, inter-line spacing of 3.7-4 Å) could be due to a bi-layer nanoribbon with successively smaller widths of each layer due to the decreasing circumference of inner nanotube shells.

Our method produced a high percentage of nanoribbons with ultra-smooth edges by simple calcination and sonication steps, which can be performed in many laboratories. The mechanism of the unzipping differs from previous methods that involved extensive solution-phase oxidation[14]. We proposed that our calcination step led to gas phase-oxidation of pre-existing defects on arc-discharge grown multiwalled carbon nanotubes. A low density structural defect was known to exist on the sidewalls and the ends of high quality arc-derived multiwalled carbon nanotubes[19]. The defects and ends were more reactive with oxygen than pristine sidewalls during 500 °C calcination, a condition used for purifying arc-discharge multiwalled carbon nanotubes without introducing new defects on sidewalls[19, 20]. Similar to oxidation of defects in the plane of graphite by oxygen [21, 22], etch pits were formed at the defects and extended from the outmost sidewall into adjacent inner walls. The depth of pits formed in this step determined the number of layers of the resulting nanoribbons. Most of our nanoribbons were single- to tri-layers, suggesting formation of etch pits through 1-3 walls on nanotubes during the 500 °C calcination step. The oxidation condition was relatively mild without creating new defects or functional groups in nanotubes, evidenced by the low Raman D-band intensity and that oxygen level measured by X-ray photoelectron spectroscopy (XPS) was similar to that of pristine multiwalled carbon nanotubes (Supplementary Fig. S5 and S6). In the solution-phase sonication process, sonochemistry and hot gas bubbles during sonication caused unzipping, which was initiated at the weak points of etch pits on nanotubes and proceeded along the tube axis. The resulting nanoribbons were separated from the inner tubes and noncovalently functionalized by PmPV via π stacking [1,23] to afford a homogeneous suspension in DCE. Scanning electron microscopy (SEM) imaging of pristine calcined and sonicated nanotubes after calcination also indicated that unzipping of nanotubes occurred during the sonication step (Supplementary Fig. S7). We carried out various control experiments (Supplementary Fig. S8 and S9) that led to an optimized unzipping protocol (see Methods). Note that our unzipping process was also applicable to CVD-grown multiwalled carbon nanotubes (Supplementary Fig. S10).

We characterized our materials by Raman spectroscopy. The Raman $I_D/I_G$ ratio is widely used to evaluate the quality of carbon nanotubes[24] and graphene materials[25]. Besides defects density and edge smoothness, $I_D/I_G$ ratio of nanoribbons is also related to the edge structures[26]. However, since the edge structures of the experimentally made nanoribbons are unknown thus far and could be random, the averaged $I_D/I_G$ may reflect the quality of nanoribbons (including edge quality, e.g., edge roughness and defects) with the same width and number of layers. The ensemble-averaged $I_D/I_G$ ratio of our final bulk product containing ~60% nanoribbons was only ~0.2 (Supplementary Fig.S5), similar to that of the starting pristine nanotubes and suggested overall low defect density in the product. We also carried out conformal Raman mapping of individual bi- and tri-layer nanoribbons deposited on SiO$_2$/Si substrates (Fig. 3A and B). The averaged $I_D/I_G$ ratio of nanoribbons with ~20 nm widths was ~0.4 (Fig. 3C), much lower than lithographic patterned nanoribbons with similar width ($I_D/I_G$~1.5, Supplementary Fig. S11) and wide nanoribbons unzipped by solution-phase oxidation ($I_D/I_G > 1$)[14, 15].

The high yield of nanoribbons suspended in an organic solution greatly simplified fabrication of nanoribbon electrical devices. We fabricated FET-like nanoribbon devices by simply making large array of source (S) and drain (D) electrodes to contact randomly deposited nanoribbons on SiO$_2$ (300 nm)/p$^{++}$-Si substrates and obtained ~15% single nanoribbon devices (Fig.4A, upper inset). The p$^{++}$-Si was used as back gate and Pd (30 nm) was used as S and D electrodes. Electrical annealing in vacuum was used[3,13] to remove adsorbates from the nanoribbons by applying a bias voltage of ~2 V. The current-gate voltage ($I_{ds}$-$V_{gs}$) curves of most nanoribbons devices showed clear Dirac points at ~0 V after electrical annealing. (Supplementary Fig. S12). Our individual nanoribbons exhibited 0.5-5$e^2$/h conductance at room temperature. The lowest resistivity (defined as $R \times W/L$, where $R$ is the resistance of the device, and $W$ and $L$ indicate nanoribbon width and channel length, respectively) at the Dirac point observed in our nanoribbons with 10-30 nm widths was 1.6 kΩ. This is the lowest resistivity of nanoribbons ever reported for nanoribbons with similar layer numbers (1-3) [1,13,14,27] (Fig. 4D, Supplementary Fig. S13). The nanoribbons exhibited mobilities up to 1500 $cm^2$/Vs for ribbons only ~14 nm in width based on gate-capacitance calculated by finite element modelling. This is the highest mobility reported for nanoribbons of similar widths[2, 27]. The lowest resistivity and highest mobility confirmed the high quality of nanoribbons produced by the new method.

Variable temperature electrical transport in nanoribbons showed that conductance of the *p*-channel of a bi-layer nanoribbon ($W$ ~14 nm, $L$ ~200 nm) increased as the device was cooled from 290 K to 50 K (Fig. 4A and B, further cooling introduced some oscillations in the $G$-$V_{gs}$ characteristics). This suggested metallic behavior for transport in the valence band of the narrow nanoribbon with reduced acoustic phonon scattering at lower temperatures. Carrier scattering in our high quality, smooth-edged nanoribbons was not dominated by defects, charged impurities or edge roughness, as in the case of nanoribbons obtained by lithographic patterning, which showed increased resistance at lower temperature due to localization effects by defects[5, 28]. At 4.2K, the conductance of our device $G$ ~3-4$e^2$/$h$ is at least one order of magnitude higher than similar previous nanoribbon devices. Conductance oscillations versus $V_{gs}$ were observed at 4.2 K, and differential conductance d$I_{ds}$/d$V_{ds}$ versus $V_{gs}$ and $V_{ds}$ exhibited interference pattern with peak conductance ~$4e^2/h$ (Fig. 4C). This was similar to Fabry-Perot interference previously observed in pristine carbon nanotubes[29], suggesting phase coherent transport and interference of several modes or subbands of electrons in the nanoribbon. Similar interference pattern was only observed in a much wider and shorter graphene sample[30]. It is remarkable that electron waves travel ~200 nm in an open-edged narrow nanoribbon ($W$ ~14 nm) without loss of phase coherence. The high conductance and phase coherent transport in the valence band of our nanoribbons again confirmed the high quality of nanoribbons made by our new approach and transparent contacts between the valance band of nanoribbons and Pd. On the other hand, the conductance of the n-channel of our nanoribbons gradually decreased at lower temperature, indicating a barrier for transport through the conduction band. This barrier is likely due to a small Schottky barrier between Pd and the conduction band of the $W$ ~14 nm nanoribbon. Band gap of the nanoribbon was estimated to be $E_g$~10-15 meV by fitting the temperature dependence of minimum conductance to thermal activation over a barrier of ~$E_g/2$ [4].

In summary, we developed a simple unzipping approach for large scale production of pristine nanoribbons from multiwalled carbon nanotubes. For the first time, narrow nanoribbons exhibiting nearly atomically smooth edges, high conductance of up to $5e^2/h$ and phase coherent transport are obtained. This simple and reliable approach makes nanoribbons easily accessible for addressing many fundamental properties predicted for these materials and for exploring their potential applications. Besides promising applications in nanoelectronics, the high quality nanoribbons also open up new avenues to control the edge chemistry of graphene nanoribbons and for the production of nanoribbon-polymer composites covalently linked at the edges.

**Methods**

**Preparation of nanoribbons**

30 mg multiwalled carbon nanotubes (Aldrich, 406074-500MG) were calcined at 500 $^{\circ}$C in a 1-inch tube furnace for 2hrs. After that, 15 mg calcined nanotubes and 7.5 mg poly (m-phenylenevinylene-co-2, 5-dioctoxy-p-phenylenevinylene) (PmPV, Aldrich, 555169-1G) were dissolved in 10 mL 1, 2-dichloroethane (DCE) and then sonicated (Cole Parmer sonicator, Model 08849-00) for 1 hr. After that, the solution was ultracentrifuged at 40,000 r.p.m (round per minute) for 2 hrs. The supernatant was collected for characterization and found to contain ~60% nanoribbons.

**Characterization of nanoribbons by AFM, TEM and Raman spectroscopy**

AFM images of nanoribbons were obtained with a Nanoscope IIIa multimode instrument in tapping mode. The samples for AFM imaging were prepared by soaking the SiO$_2$/Si substrates in the nanoribbons suspension for 15 min, rinsing with isopropanol and then blowing dried. Before AFM imaging, the substrates were calcined at 350 $^{\circ}$C for 20 min to remove PmPV.

We characterized our nanoribbons using a FEI Tecnai G2 F20 X-TWIN TEM at an accelerating voltage of 120 kV or 200 kV. The TEM samples were prepared by soaking porous Si grids (SPI Supplies, US200-P15Q UltraSM 15nm Porous TEM Windows) in a nanoribbons suspension overnight and then calcined at 400 $^{\circ}$C for 20 min.

For characterization of individual nanoribbons by Raman spectroscopy, we obtained low density nanoribbons on $SiO_2$/Si substrates with makers by soaking the substrates in nanoribbons suspensions for 2 min. Then we located individual nanoribbons with markers by AFM. Raman spectra of individual nanoribbons were collected with Horiba Jobin Yvon LabRAM HR Raman microscope with a 633 nm He-Ne laser excitation (spot size ~1 μm, power ~10 mW). The step size of mapping was 100 nm and the integration time was 5 s at each spot.

**Fabrication of nanoribbons devices**

We used electron-beam lithography followed by electron-beam evaporation of palladium (30 nm) to fabricate a large array of 98 source- and drain-electrodes on 300-nm $SiO_2$ /$p^{++}$ Si substrates with pre-deposited nanoribbons. The channel length of these devices was ~250 nm and the width of source and drain electrodes was ~5 μm. The devices were then annealed in Ar at 220 ºC for 15 min to improve the contact quality. AFM was then used to identify devices with a single nanoribbon connection. The yield of such devices on a chip is ~10-15%.

**Reference**

1. Li, X. L. *et al*. Chemically derived, ultrasmooth graphene nanoribbon semiconductors. *Science* **319**, 1229-1232 (2008).

2. Wang, X. R. *et al.* Room-temperature all-semiconducting sub-10-nm graphene nanoribbon field-effect transistors. *Phys. Rev. Lett.* **100**, 206803 (2008).

3. Wang, X. R. *et al*. N-doping of graphene Through electrothermal reactions with ammonia. *Science* **324**, 768-771 (2009).

4. Chen, Z. H., Lin, Y. M., Rooks, M. J.&Avouris, P. Graphene nano-ribbon electronics. *Physica. E* **40**, 228-232 (2007).

5. Han, M. Y., Ozyilmaz, B., Zhang, Y. B.&Kim, P. Energy band-gap engineering of graphene nanoribbons. *Phys. Rev. Lett.* **98**, 206805 (2007).

6. Cresti, A. *et al*. Charge transport in disordered graphene-based low dimensional materials. *Nano Res*. **1**, 361-394 (2008).

7. Tapaszto, L., Dobrik, G., Lambin, P.&Biro, L. P. Tailoring the atomic structure of graphene nanoribbons by scanning tunnelling microscope lithography. *Nat. Nanotechnol*. **3**, 397-401 (2008).

8. Datta, S. S., Strachan, D. R., Khamis, S. M.&Johnson, A. T. C. Crystallographic etching of few-layer graphene. *Nano Lett*. **8**, 1912-1915 (2008).

9. Ci, L.J. *et al*. Controlled nanocutting of graphene. *Nano Res*. **1**, 116-122 (2008).

10. Campos, L. C., Manfrinato, V. R., Sanchez-Yamagishi, J. D., Kong, J. & Jarillo-Herrero, P. Anisotropic Etching and nanoribbon formation in single-layer graphene. *Nano Lett*. **9**, 2600-2604 (2009).

11. Campos-Delgado, J. *et al*. Bulk production of a new form of $sp^2$ carbon: Crystalline graphene nanoribbons. *Nano Lett*. **8**, 2773-2778 (2008).

12. Wu, Z.S. *et al*. Efficient synthesis of graphene nanoribbons sonochemically cut from graphene sheets. *Nano Res*. **3**, 16-22 (2010).

13. Jiao, L. Y., Zhang, L., Wang, X. R., Diankov, G. & Dai, H. J. Narrow graphene nanoribbons from carbon nanotubes. *Nature* **458**, 877-880 (2009).

14. Kosynkin, D. V. *et al*. Longitudinal unzipping of carbon nanotubes to form graphene nanoribbons. *Nature* **458**, 872-876 (2009).

15. Zhang, Z. X., Sun, Z. Z., Yao, J., Kosynkin, D. V. & Tour, J. M. Transforming carbon nanotube devices into nanoribbon devices. *J. Am. Chem. Soc*. **131**, 13460-13463 (2009).

16. Cano-Marquez, A. G. *et al*. Ex-MWNTs: Graphene sheets and ribbons produced by lithium intercalation and exfoliation of carbon nanotubes. *Nano Lett*. **9**, 1527-1533 (2009).

17. Elías, A. L., *et al*. Longitudinal cutting of pure and doped carbon nanotubes to form graphitic nanoribbons using metal clusters as nanoscalpels. *Nano Lett*. **10**, 366-372 (2010).

18. Kim, W. S. *et al*. Fabrication of graphene layers from multiwalled carbon nanotubes using high dc pulse. *Appl. Phys. Lett*. **95**, 083103 (2009).

19. Colbert, D.T. *et al.* Growth and sintering of fullerene nanotubes. *Science* 266 (5188), 1218-1222 (1994).

20. Barinov, A., Gregoratti, L., Dudin, P., La Rosa, S. & Kiskinova, M. Imaging and Spectroscopy of multiwalled carbon nanotubes during oxidation: defects and oxygen bonding. *Adv. Mater.* **21**, 1916-1920 (2009).

21. Stevens, F., Kolodny, L. A. & Beebe, T. P. Kinetics of graphite oxidation: Monolayer and multilayer etch pits in HOPG studied by STM. *J. Phys. Chem. B* **102**, 10799-10804 (1998).

22. Lee, S. M. *et al*. Defected-induced oxidation of graphite. *Phys. Rev. Lett*. **82**, 217 (1999).

23. Chen, R.J., Zhang, Y.G., Wang, D.W., & Dai, H.J., Noncovalent sidewall functionalization of single-walled carbon nanotubes for protein immobilization. *J. Am. Chem. Soc.* **123**, 3838-3839 (2001).

24. Dresselhaus, M. S., Dresselhaus, G., Saito, R. & Jorio, A. Raman spectroscopy of carbon nanotubes. *Phys. Rep.* **409**, 47-99 (2005).

25. Ni, Z.H., Wang, Y.Y., Yu, T. & Shen, Z.X. Raman spectroscopy and imaging of graphene. *Nano Res.* **1**, 273-291 (2008).

26. Gupta, A.K, Russin, T. J., Gutiérrez, H. R.& Eklund, P.C. Probing graphene edges *via* Raman scattering. *ACS Nano* **3**, 45-52 (2009).

27. Lin, Y. M.& Avouris, P. Strong suppression of electrical noise in bilayer graphene nanodevices. *Nano Lett.* **8**, 2119–2125 (2008).

28. Han, M. Y., Brant, J. C., & Kim, P. Electron transport in disordered graphene nanoribbons. arXiv:0910.4808v1 (2009).

29. Liang, W. *et al*. Fabry-Perot interference in a nanotube electronwaveguide. *Nature* **411**, 665–669 (2001).

30. Todd, K., Chou, H.T., Amasha, S., & Goldhaber-Gordon, D. Quantum dot behavior in graphene nanoconstrictions. *Nano Lett* **9**, 416-421 (2009).

**Acknowledgements** This work was supported by MARCO-MSD, Intel and ONR.

**Author contributions**

H.D. and L.J. conceived and designed the experiments. L.J., X.W., G.D. and H.W. performed the experiments and analyzed the data. H.D. and L.J. co-wrote the manuscript. All authors discussed the results and commented on the manuscript.

**Author Information**

**Legend**

**Figure 1 Unzipping of nanotubes by a new two step method in gas and liquid phases. (A)** Schematic of the unzipping processes. In the mild gas-phase oxidation step, oxygen reacted with pre-existed defects on nanotubes to form etch pits on the sidewalls. In the solution-phase sonication step, sonochemistry and hot gas bubbles enlarged the pits and unzipped the tubes. **(B)** to **(D)**, AFM images of pristine, partially and fully unzipped nanotubes, respectively. The heights of nanoribbons shown in **(C)** and **(D)** are 1.4 and 1.6 nm, respectively, much lower than the pristine nanotube shown in **(B)** (height ~9 nm).

**Figure 2 Microscopy imaging of nanoribbons. (A)** An AFM image of unzipped nanotubes deposited on $SiO_2$/Si substrate, showing a high percentage of single-, bi- and tri-layer nanoribbons (~60%). **(B)** A zoom-in AFM image of a part in **(A)**, showing smooth edges of nanoribbons. The heights and widths of the three nanoribbons from top to bottom were: 1.8 nm, 18 nm; 1.4 nm, 48 nm; 1.4 nm, 22 nm, respectively. **(C)** A TEM (acceleration voltage= 200 kV) image of a ~12-nm-wide nanoribbon with a kink due to folding. The dark spots on the substrate are nanocrystalline domains within the porous silicon grids. **(D)** TEM image of the kink on the nanoribbon shown in **(C)**. **(E)** and **(F)** TEM (acceleration voltage = 120 kV) images of nanoribbons suspended over the holes of porous silicon grids, showing nearly atomically smooth edges. The widths of the nanoribbons shown in **(E)** and **(F)** were ~12 and 10 nm, respectively. The amorphous coating on the nanoribbon shown in **(E)** was PmPV used to suspend nanoribbons.

**Figure 3 Raman spectroscopy of nanoribbons. (A)** and **(B)**, Raman spectrum of a bi- (height ~1.5 nm) and tri-layer (height ~1.8nm) nanoribbon ($W$ ~20 nm) on $SiO_2$/Si

substrates, respectively. Inset, AFM images and G-band images of the same nanoribbons on the same length scale. The $I_D/I_G$ ratios of these two nanoribbons are 0.3 and 0.5, respectively. **(C)** Comparison of averaged $I_D/I_G$ of 5-10 bi-layer nanoribbons with ~20 nm widths made by different methods, including method present in this paper, lithographic patterning (Supplementary Fig. S11) and plasma unzipping[13].

**Figure 4 Electrical transport measurements of nanoribbons.** **(A)** $G$-$V_{gs}$ curves of a 14-nm-wide bi-layer nanoribbon at 20 K, 100 K and 290 K, $V_{ds}$= 1 mV. Upper inset, schematic of nanoribbons devices made by randomly contacting. Lower inset, AFM image of this nanoribbon device. **(B)** $G$-$T$ relationship of the nanoribbon shown in **(A)** at $V_{gs}$ of -30 V. The conductance increased as cooled from room temperature to 50 K. **(C)** Top panel: Differential conductance $dI_{ds}/dV_{ds}$ versus $V_{ds}$ and $V_{gs}$ of the nanoribbon shown in **(A)** measured in a cryogenic insert at 4.2 K shows a Fabry-Perot-like interference pattern. Bottom panel: the $dI_{ds}/dV_{ds}$ versus $V_{gs}$ curve of the nanoribbon shown in **(A)** at $V_{ds}$= 0 mV shows conductance peaks and valleys. **(D)** Comparison of room temperature resistivity of bi-layer nanoribbons with 10-30 nm widths made by different methods, including method present in this paper, lithographic patterning[27] (Supplementary Fig. S13), sonochemical method[1] and plasma unzipping[13]. The resistivity (~1 MΩ) of wide nanoribbons made by unzipping nanotubes in solution-phase[14, 15] was not included in the comparison.

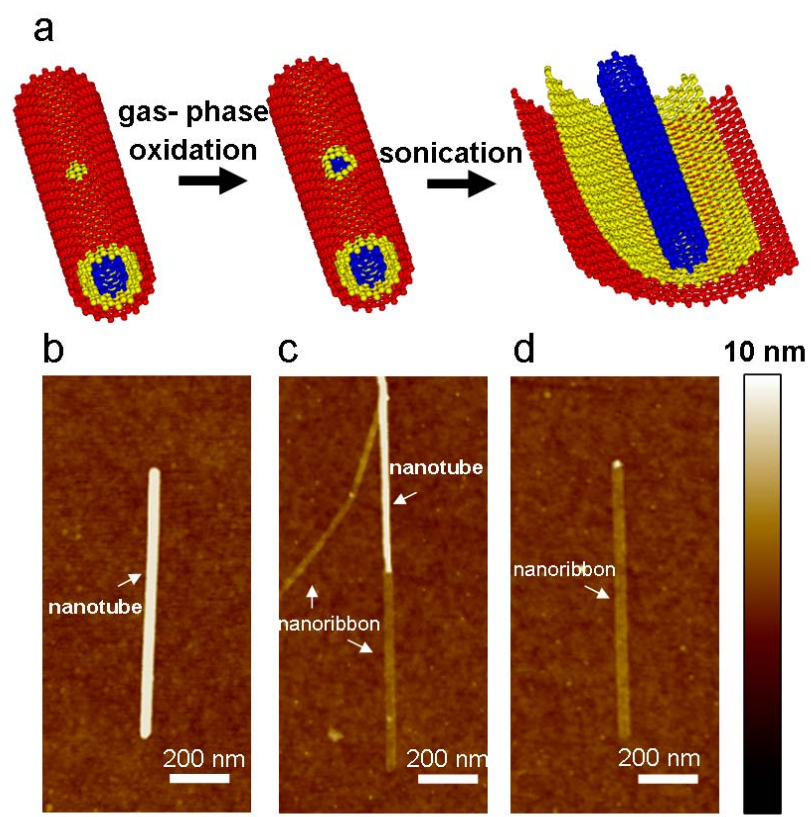

**Figure 1**

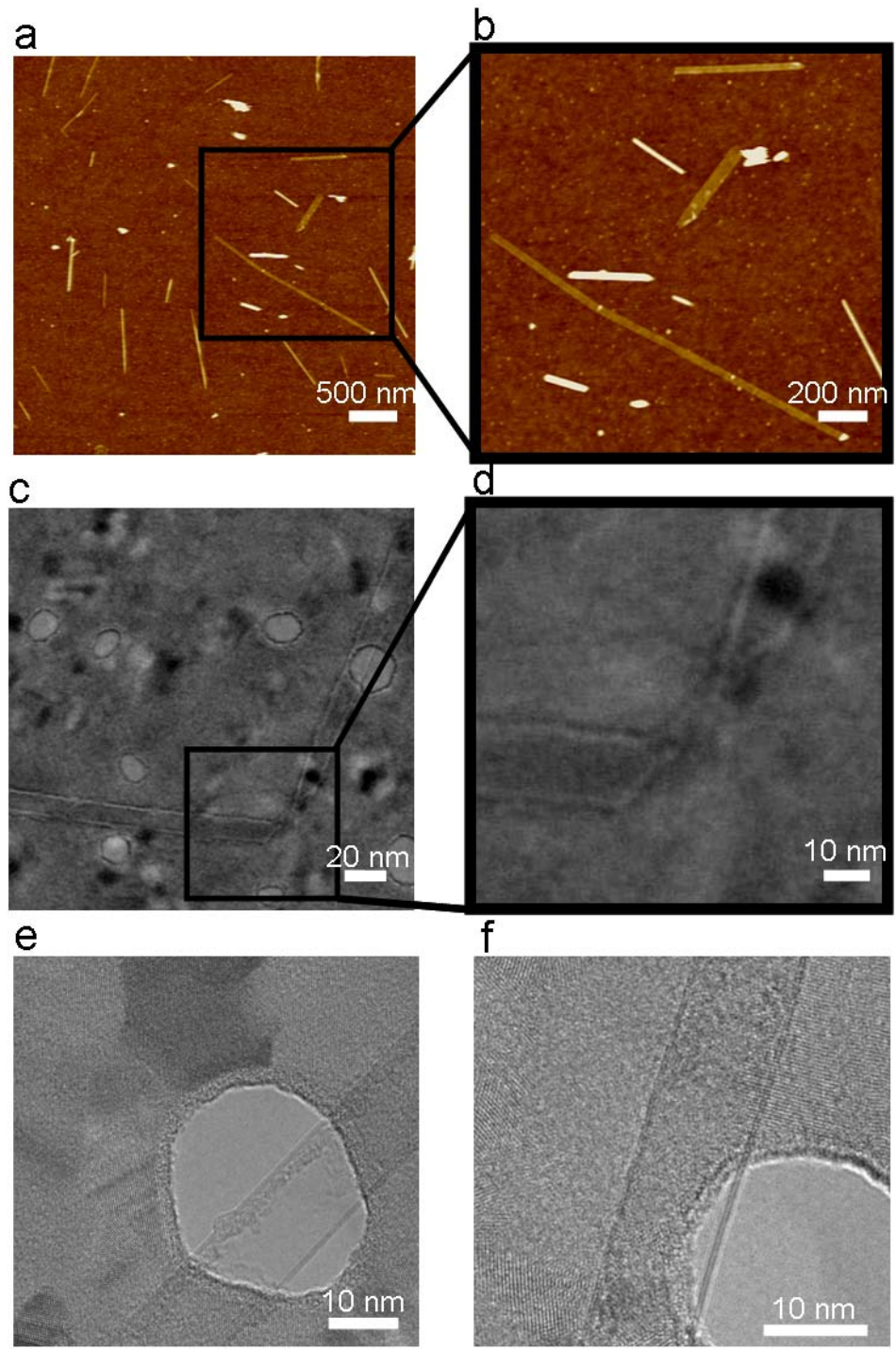

**Figure 2**

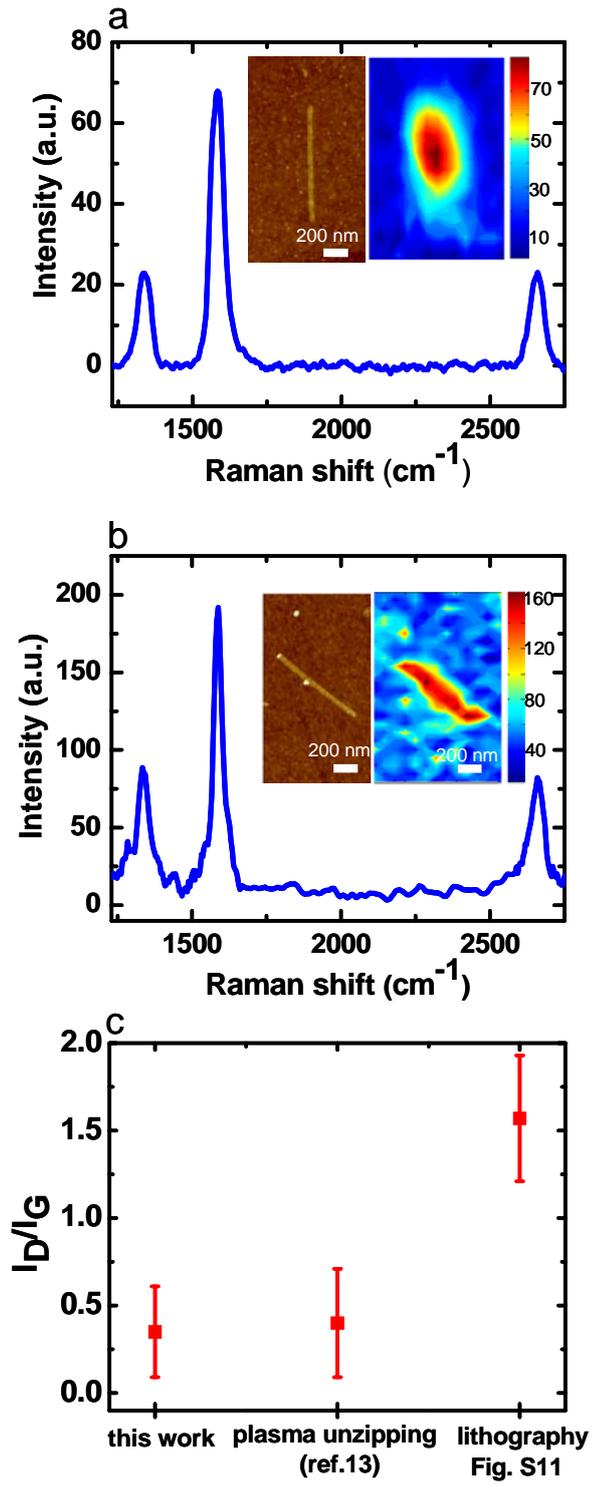

**Figure 3**

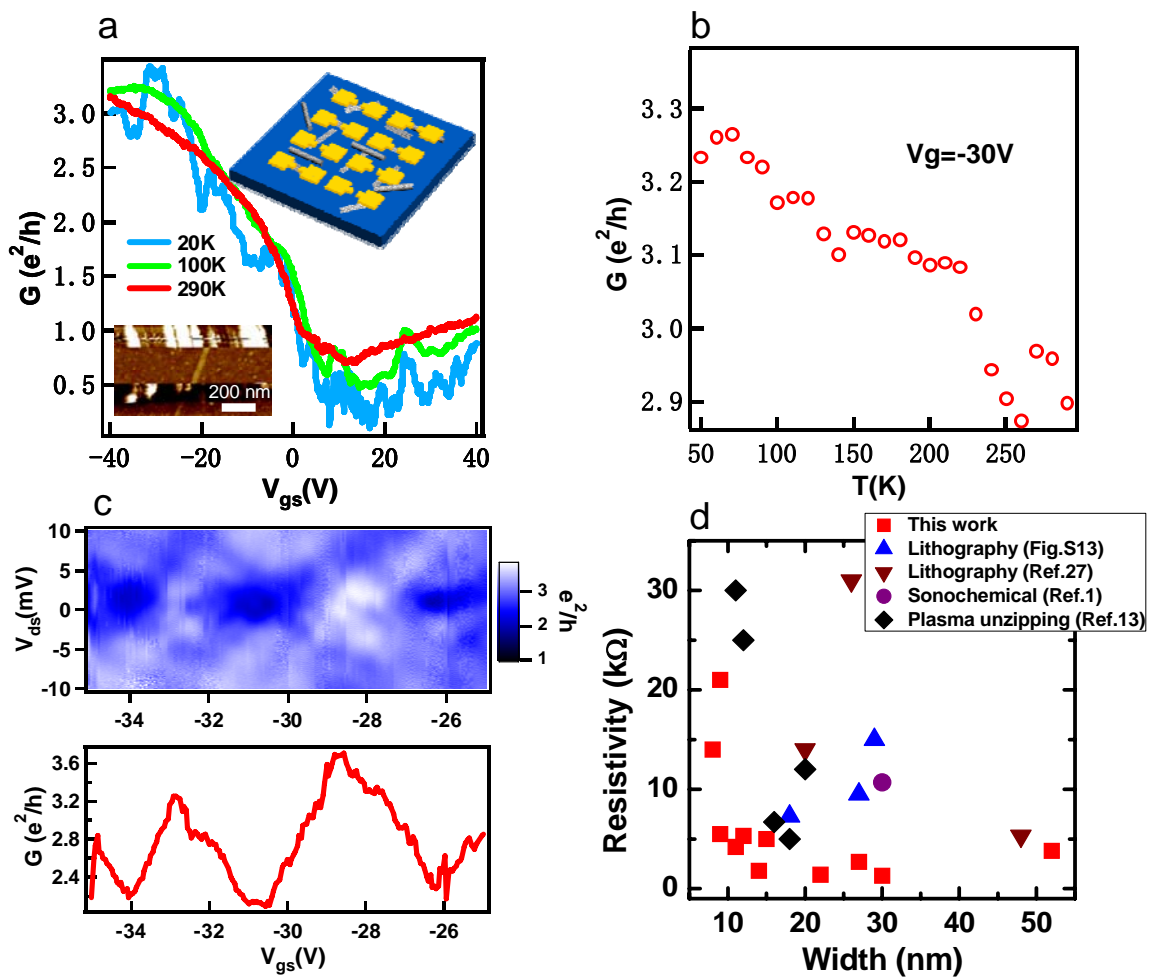

**Figure 4**